\documentclass[epj,nopacs]{svjoura}
\usepackage{amsmath}
\usepackage{graphics}
\usepackage{graphicx}
\usepackage{epsfig}
\usepackage{amssymb}
\usepackage{units}
\usepackage{cite}
\usepackage{calc}
\usepackage{endnotes}

\newcommand{\dndeta}[1]{${\rm d}N_{\rm ch}/{\rm d}\eta = #1$}
\newcommand{\dNdEta}{${\rm d}N_{\rm ch}/{\rm d}\eta$ }
\newcommand{\tblspc}{\rule{0pt}{2.5ex}}
\newcounter{vers}

\usepackage{preprintcover}
\PreprintIdNumber{}
\PreprintCoverPaperTitle{First proton--proton collisions at the LHC as observed with the ALICE detector: measurement of the charged-particle pseudorapidity density at ${\sqrt{{s}}= 900}$\,GeV}
\PreprintCoverAbstract{On 23$^\mathrm{rd}$ November 2009, during the early commissioning  of the CERN Large Hadron Collider (LHC), two counter-rotating proton bunches were circulated for the first time concurrently in the machine, at the LHC injection energy of 450~GeV per beam. Although the proton intensity was very low, with only one pilot bunch per beam, and no systematic attempt was made to optimize the collision optics, all LHC experiments reported a number of  collision candidates. In the ALICE experiment, the collision region was centred very well in both the longitudinal and transverse directions and 284 events were recorded in coincidence with the two passing proton bunches. The events were immediately reconstructed and analyzed both online and offline.
We have used these events to measure the  pseudorapidity density of
charged primary particles in the central region. 
In the range $|\eta|<0.5$, we obtain \dndeta{3.10 \pm 0.13 {\mathrm(stat.)} \pm 0.22 {\mathrm(syst.)}} for all inelastic interactions, and \dndeta{3.51 \pm 0.15 {\mathrm(stat.)} \pm 0.25 {\mathrm(syst.)}} for non-single diffractive
interactions. These results are consistent with previous measurements in
proton--antiproton interactions at the same centre-of-mass energy at
the CERN Sp$\overline{\rm p}$S collider. They also illustrate the excellent
functioning and rapid progress of the LHC accelerator, and of both the hardware
and software of the ALICE experiment, in this early start-up phase.}
\PreprintJournalName{EPJC}

\begin{document}
\hugehead
\title{First proton--proton collisions at the LHC as observed with the ALICE detector: measurement of the charged-particle pseudorapidity density at ${\sqrt{{s}}= 900}$\,GeV}
\subtitle{ALICE collaboration}
\author{
K.~Aamodt\inst{78} \and
N.~Abel\inst{43} \and
U.~Abeysekara\inst{30} \and
A.~Abrahantes~Quintana\inst{42} \and
A.~Acero\inst{63} \and
D.~Adamov\'{a}\inst{86} \and
M.M.~Aggarwal\inst{25} \and
G.~Aglieri~Rinella\inst{40} \and
A.G.~Agocs\inst{18} \and
S.~Aguilar~Salazar\inst{66}
% ...
\renewcommand{\notesname}{Affiliation notes}
\endnotetext[1]{Also at\inst{26}}
\endnotetext[2]{Also at\inst{36}}
\endnotetext[3]{Now at\inst{80}}
\endnotetext[4]{Now at\inst{32}}
\endnotetext[5]{Now at\inst{37}}
\endnotetext[6]{Now at\inst{22}}
\endnotetext[7]{Now at\inst{17}}
\endnotetext[8]{Also at\inst{46}}
\endnotetext[9]{Now at\inst{57}}
\endnotetext[10]{Now at\inst{40}}
\endnotetext[11]{Now at\inst{49}}
\endnotetext[12]{Now at\inst{99}}
\endnotetext[13]{Now at\inst{6}}
\endnotetext[14]{Now at\inst{72}}
\endnotetext[15]{Now at: University of Technology and Austrian Academy of Sciences, Vienna, Austria}
\endnotetext[16]{Also at\inst{60}}
\endnotetext[17]{Also at\inst{40}}
\endnotetext[18]{Now at: Secci\'{o}n F\'{\i}sica, Departamentode de Ciencias, Pontificia Universidad Cat\'{o}lica del Per\'{u}, Lima, Peru}
\endnotetext[19]{Deceased}
\endnotetext[20]{Now at\inst{74}}
\endnotetext[21]{Now at\inst{105}}
\endnotetext[22]{Also at: Centro Fermi -- Centro Studi e Ricerche e Museo Storico della Fisica ``Enrico Fermi'', Rome, Italy}
\endnotetext[23]{Now at\inst{5}}
\endnotetext[24]{Also at\inst{41}}
\endnotetext[25]{Now at\inst{101}}
\endnotetext[26]{Now at\inst{30}}
\endnotetext[27]{Now at\inst{89}}
\endnotetext[28]{Also at\inst{78}}
\endnotetext[29]{Now at\inst{44}}
\endnotetext[30]{Now at\inst{33}}
\endnotetext[31]{Now at\inst{8}}
\endnotetext[32]{Now at\inst{4}}
\endnotetext[33]{Also at\inst{72}}
\endnotetext[34]{Now at\inst{73}}
\endnotetext[35]{Now at\inst{62}}
\endnotetext[36]{Now at: Centre de Calcul IN2P3, Lyon, France}
\endnotetext[37]{Now at\inst{55}}
\endnotetext[38]{Also at\inst{79}}
\endnotetext[39]{Also at\inst{14}}
\bigskip
\theendnotes
\section*{Collaboration institutes}
}

\institute{
Dipartimento di Scienze e Tecnologie Avanzate dell'Universit\`{a} del Piemonte Orientale and Gruppo Collegato INFN, Alessandria, Italy
\and
Department of Physics Aligarh Muslim University, Aligarh, India
\and
National Institute for Nuclear and High Energy Physics (NIKHEF), Amsterdam, Netherlands \and
Physics Department, University of Athens, Athens, Greece
\and
Dipartimento Interateneo di Fisica `M.~Merlin' and Sezione INFN, Bari, Italy
\and
Sezione INFN, Bari, Italy
\and
China Institute of Atomic Energy, Beijing, China
\and
Department of Physics and Technology, University of Bergen, Bergen, Norway
\and
Faculty of Engineering, Bergen University College, Bergen, Norway
\and
Lawrence Berkeley National Laboratory, Berkeley, California, United States
\and
Institute of Physics, Bhubaneswar, India
\and
School of Physics and Astronomy, University of Birmingham, Birmingham, United Kingdom
\and
Dipartimento di Fisica dell'Universit\`{a} and Sezione INFN, Bologna, Italy
\and
Sezione INFN, Bologna, Italy
\and
Faculty of Mathematics, Physics and Informatics, Comenius University, Bratislava, Slovakia
\and
Institute of Space Sciences (ISS), Bucharest, Romania
\and
National Institute for Physics and Nuclear Engineering, Bucharest, Romania
\and
KFKI Research Institute for Particle and Nuclear Physics, Hungarian Academy of Sciences, Budapest, Hungary
\and
Dipartimento di Fisica dell'Universit\`{a} and Sezione INFN, Cagliari, Italy
\and
Sezione INFN, Cagliari, Italy
\and
Universidade Estadual de Campinas (UNICAMP), Campinas, Brazil
\and
Physics Department, University of Cape Town, iThemba Laboratories, Cape Town, South Africa
\and
Dipartimento di Fisica e Astronomia dell'Universit\`{a} and Sezione INFN, Catania, Italy
\and
Sezione INFN, Catania, Italy
\and
Physics Department, Panjab University, Chandigarh, India
\and
Laboratoire de Physique Corpusculaire (LPC), Clermont Universit\'{e}, Universit\'{e} Blaise Pascal, CNRS--IN2P3, Clermont-Ferrand, France
\and
Department of Physics, Ohio State University, Columbus, Ohio, United States
\and
Niels Bohr Institute, University of Copenhagen, Copenhagen, Denmark
\and
The Henryk Niewodniczanski Institute of Nuclear Physics, Polish Academy of Sciences, Cracow, Poland
\and
Physics Department, Creighton University, Omaha, Nebraska, United States
\and
Universidad Aut\'{o}noma de Sinaloa, Culiac\'{a}n, Mexico
\and
ExtreMe Matter Institute EMMI, GSI Helmholtzzentrum f\"{u}r Schwerionenforschung, Darmstadt, Germany
\and
Institut f\"{u}r Kernphysik, Technische Universit\"{a}t Darmstadt, Darmstadt, Germany
\and
Wayne State University, Detroit, Michigan, United States
\and
Joint Institute for Nuclear Research (JINR), Dubna, Russia
\and
Frankfurt Institute for Advanced Studies, Johann Wolfgang Goethe-Universit\"{a}t Frankfurt, Frankfurt, Germany
\and
Institut f\"{u}r Kernphysik, Johann Wolfgang Goethe-Universit\"{a}t Frankfurt, Frankfurt, Germany
\and
Laboratori Nazionali di Frascati, INFN, Frascati, Italy
\and
Petersburg Nuclear Physics Institute, Gatchina, Russia
\and
European Organization for Nuclear Research (CERN), Geneva, Switzerland
\and
Laboratoire de Physique Subatomique et de Cosmologie (LPSC), Universit\'{e} Joseph Fourier, CNRS-IN2P3, Institut Polytechnique de Grenoble, Grenoble, France
\and
Centro de Aplicaciones Tecnol\'{o}gicas y Desarrollo Nuclear (CEADEN), Havana, Cuba
\and
Kirchhoff-Institut f\"{u}r Physik, Ruprecht-Karls-Universit\"{a}t Heidelberg, Heidelberg, Germany
\and
Physikalisches Institut, Ruprecht-Karls-Universit\"{a}t Heidelberg, Heidelberg, Germany
\and
Hiroshima University, Hiroshima, Japan
\and
University of Houston, Houston, Texas, United States
\and
Physics Department, University of Rajasthan, Jaipur, India
\and
Physics Department, University of Jammu, Jammu, India
\and
Helsinki Institute of Physics (HIP) and University of Jyv\"{a}skyl\"{a}, Jyv\"{a}skyl\"{a}, Finland
\and
Kangnung National University, Kangnung, South Korea
\and
Scientific Research Technological Institute of Instrument Engineering, Kharkov, Ukraine
\and
Bogolyubov Institute for Theoretical Physics, Kiev, Ukraine
\and
University of Tennessee, Knoxville, Tennessee, United States
\and
Saha Institute of Nuclear Physics, Kolkata, India
\and
Variable Energy Cyclotron Centre, Kolkata, India
\and
Fachhochschule K\"{o}ln, K\"{o}ln, Germany
\and
Faculty of Science, P.J.~\v{S}af\'{a}rik University, Ko\v{s}ice, Slovakia
\and
Institute of Experimental Physics, Slovak Academy of Sciences, Ko\v{s}ice, Slovakia
\and
Laboratori Nazionali di Legnaro, INFN, Legnaro, Italy
\and
Lawrence Livermore National Laboratory, Livermore, California, United States
\and
Division of Experimental High Energy Physics, University of Lund, Lund, Sweden
\and
Universit\'{e} de Lyon 1, CNRS/IN2P3, Institut de Physique Nucl\'{e}aire de Lyon, Lyon, France
\and
Centro de Investigaciones Energ\'{e}ticas Medioambientales y Tecnol\'{o}gicas (CIEMAT), Madrid, Spain
\and
Centro de Investigaci\'{o}n y de Estudios Avanzados (CINVESTAV), Mexico City and M\'{e}rida, Mexico
\and
Instituto de Ciencias Nucleares, Universidad Nacional Aut\'{o}noma de M\'{e}xico, Mexico City, Mexico
\and
Instituto de F\'{\i}sica, Universidad Nacional Aut\'{o}noma de M\'{e}xico, Mexico City, Mexico
\and
Institute for Nuclear Research, Academy of Sciences, Moscow, Russia
\and
Institute for Theoretical and Experimental Physics, Moscow, Russia
\and
Moscow Engineering Physics Institute, Moscow, Russia
\and
Russian Research Centre Kurchatov Institute, Moscow, Russia
\and
Indian Institute of Technology, Mumbai, India
\and
Institut f\"{u}r Kernphysik, Westf\"{a}lische Wilhelms-Universit\"{a}t M\"{u}nster, M\"{u}nster, Germany
\and
SUBATECH, Ecole des Mines de Nantes, Universit\'{e} de Nantes, CNRS-IN2P3, Nantes, France
\and
Yale University, New Haven, Connecticut, United States
\and
Budker Institute for Nuclear Physics, Novosibirsk, Russia
\and
Oak Ridge National Laboratory, Oak Ridge, Tennessee, United States
\and
Institut de Physique Nucl\'{e}aire d'Orsay (IPNO), Universit\'{e} Paris-Sud, CNRS-IN2P3, Orsay, France
\and
Department of Physics, University of Oslo, Oslo, Norway
\and
Dipartimento di Fisica dell'Universit\`{a} and Sezione INFN, Padova, Italy
\and
Sezione INFN, Padova, Italy
\and
Faculty of Nuclear Sciences and Physical Engineering, Czech Technical University in Prague, Prague, Czech Republic
\and
Institute of Physics, Academy of Sciences of the Czech Republic, Prague, Czech Republic
\and
Institute for High Energy Physics, Protvino, Russia
\and
Benem\'{e}rita Universidad Aut\'{o}noma de Puebla, Puebla, Mexico
\and
Pusan National University, Pusan, South Korea
\and
Nuclear Physics Institute, Academy of Sciences of the Czech Republic, \v{R}e\v{z} u Prahy, Czech Republic
\and
Dipartimento di Fisica dell'Universit\`{a} `La Sapienza' and Sezione INFN, Rome, Italy
\and
Sezione INFN, Rome, Italy
\and
Commissariat \`{a} l'Energie Atomique, IRFU, Saclay, France
\and
Dipartimento di Fisica `E.R.~Caianiello' dell'Universit\`{a} and Sezione INFN, Salerno, Italy
\and
California Polytechnic State University, San Luis Obispo, California, United States
\and
Departamento de F\'{\i}sica de Part\'{\i}culas and IGFAE, Universidad de Santiago de Compostela, Santiago de Compostela, Spain
\and
Universidade de S\~{a}o Paulo (USP), S\~{a}o Paulo, Brazil
\and
Russian Federal Nuclear Center (VNIIEF), Sarov, Russia
\and
Department of Physics, Sejong University, Seoul, South Korea
\and
Yonsei University, Seoul, South Korea
\and
Technical University of Split FESB, Split, Croatia
\and
V.~Fock Institute for Physics, St. Petersburg State University, St. Petersburg, Russia
\and
Institut Pluridisciplinaire Hubert Curien (IPHC), Universit\'{e} de Strasbourg, CNRS-IN2P3, Strasbourg, France
\and
University of Tokyo, Tokyo, Japan
\and
Dipartimento di Fisica Sperimentale dell'Universit\`{a} and Sezione INFN, Turin, Italy \and
Sezione INFN, Turin, Italy
\and
Dipartimento di Fisica dell'Universit\`{a} and Sezione INFN, Trieste, Italy
\and
Sezione INFN, Trieste, Italy
\and
University of Tsukuba, Tsukuba, Japan
\and
Institute for Subatomic Physics, Utrecht University, Utrecht, Netherlands
\and
Soltan Institute for Nuclear Studies, Warsaw, Poland
\and
Warsaw University of Technology, Warsaw, Poland
\and
Purdue University, West Lafayette, Indiana, United States
\and
Zentrum f\"{u}r Technologietransfer und Telekommunikation (ZTT), Fachhochschule Worms, Worms, Germany
\and
Hua-Zhong Normal University, Wuhan, China
\and
Yerevan Physics Institute, Yerevan, Armenia
\and
Rudjer Bo\v{s}kovi\'{c} Institute, Zagreb, Croatia
} 
%
%
%
\date{}
%

\abstract{On 23$^\mathrm{rd}$ November 2009, during the early commissioning  of the CERN Large Hadron Collider (LHC), two counter-rotating proton bunches were circulated for the first time concurrently in the machine, at the LHC injection energy of 450~GeV per beam. Although the proton intensity was very low, with only one pilot bunch per beam, and no systematic attempt was made to optimize the collision optics, all LHC experiments reported a number of  collision candidates. In the ALICE experiment, the collision region was centred very well in both the longitudinal and transverse directions and 284 events were recorded in coincidence with the two passing proton bunches. The events were immediately reconstructed and analyzed both online and offline.
We have used these events to measure the  pseudorapidity density of
charged primary particles in the central region. 
In the range $|\eta|<0.5$, we obtain \dndeta{3.10 \pm 0.13 {\mathrm(stat.)} \pm 0.22 {\mathrm(syst.)}} for all inelastic interactions, and \dndeta{3.51 \pm 0.15 {\mathrm(stat.)} \pm 0.25 {\mathrm(syst.)}} for non-single diffractive
interactions. These results are consistent with previous measurements in
proton--antiproton interactions at the same centre-of-mass energy at
the CERN Sp$\overline{\rm p}$S collider. They also illustrate the excellent
functioning and rapid progress of the LHC accelerator, and of both the hardware
and software of the ALICE experiment, in this early start-up phase.} 


%
\maketitle
%
\section{Introduction}

The very first proton--proton collisions at Point 2 of the CERN Large
Hadron Collider (LHC)~\cite{LHC} occurred in the afternoon of 23$^\mathrm {rd}$ November 2009, at a
centre-of-mass energy $\sqrt{s} = \unit[900]{GeV}$, during the
commissioning of the accelerator. This publication, based on 284 events
recorded in the ALICE detector~\cite{ALICEdet} on that day,
describes a determination of the
pseudorapidity density of charged primary particles\footnote{Here,
primary particles are defined as prompt particles produced in the collision and
all decay products, except
products from weak decays of strange particles such as K$^0_{\rm s}$ and $\Lambda$.}
${\rm d}N_{\rm ch}/{\rm d}\eta$ ($\eta \equiv - \ln \tan \theta / 2$, where $\theta$ is the polar angle with respect to the beam line) in the central pseudorapidity region. The purpose of this study is to compare with previous measurements for proton--antiproton (p$\overline{\rm p}$) collisions at the same energy~\cite{UA5_dNdEta}, and to establish a reference for comparison with forthcoming measurements at higher LHC energies.

The event sample collected with our trigger contains three different classes of inelastic interactions, i.e. collisions where new particles are produced: non-diffractive, single-diffractive, and double-diffractive\footnote{Inelastic pp collisions are usually divided into these classes depending on the fate of the interacting protons. If one (both) incoming beam particle(s) are excited into a high-mass state, the process is called single (double) diffraction; otherwise the events are classified as non-diffractive. Particles emitted in diffractive reactions are usually found at rapidities close to that of the parent proton.}. Experimentally we cannot distinguish between these classes, which, however, are selected by our trigger with different efficiencies\footnote{We estimate the \emph{trigger efficiency} for each class using the process-type information provided by Monte Carlo generators; the values vary by up to a factor of two between classes and are listed in Section 3. The \emph{relative abundance} of each class is taken from published data (see text).}.

In order to compare our data with those of other experiments, we provide the result with two different normalizations: the first one (INEL) corresponds to the sum of all inelastic interactions and corrects the trigger bias individually for all event classes, by weighting them, each with its own estimated trigger efficiency and abundance. The second normalization (non-single-diffractive or NSD) applies this correction for non-diffractive and double-diffractive processes only, while removing, on average, the single-diffractive contribution.

Multiparticle production is rather successfully describ\-ed by phenomenological models with Pomeron exchange, which dominates at high energies~\cite{QGSM,DPM}. These models relate the energy dependence of the total cross section to that of the multiplicity production using a small number of parameters, and are the basis for several Monte Carlo event generators describing soft hadron collisions (see for example~\cite{QGSMMC,DPMJet,PHOJET}).
According to these models, it is expected that the charged-particle density
increases by a factor $1.7$ and $1.9$ when raising the LHC centre-of-mass
energy from $\unit[900]{GeV}$ to $7$ and $\unit[14]{TeV}$ respectively
(i.e. intermediate and nominal LHC energies). The difference in
charged-particle densities between p$\overline{\rm p}$ and pp interactions
is predicted to decrease as $1 / \sqrt{s}$ at high
energies~\cite{difpap}. This difference was last measured at the CERN ISR
to be in the range $1.5$--$3$\,\%~\cite{R210} at $\sqrt{s} =
\unit[53]{GeV}$. Extrapolating these values to $\sqrt{s} =
\unit[900]{GeV}$, one obtains a very small difference of about $0.1$--$0.2$\,\%.
Therefore, we will compare our measurement to existing p$\overline{\rm p}$
data and also to different Monte Carlo models.

This article is organized as follows: Section 2 describes the experimental conditions during data taking; the main features of the ALICE detector subsystems used for this analysis are decribed in Section 3; Section 4 is dedicated to the event selection and data analysis; the results are discussed in Section 5 and Section 6 contains the conclusion.

\begin{figure*}[p]
\centering
\includegraphics[width=0.95\textwidth]{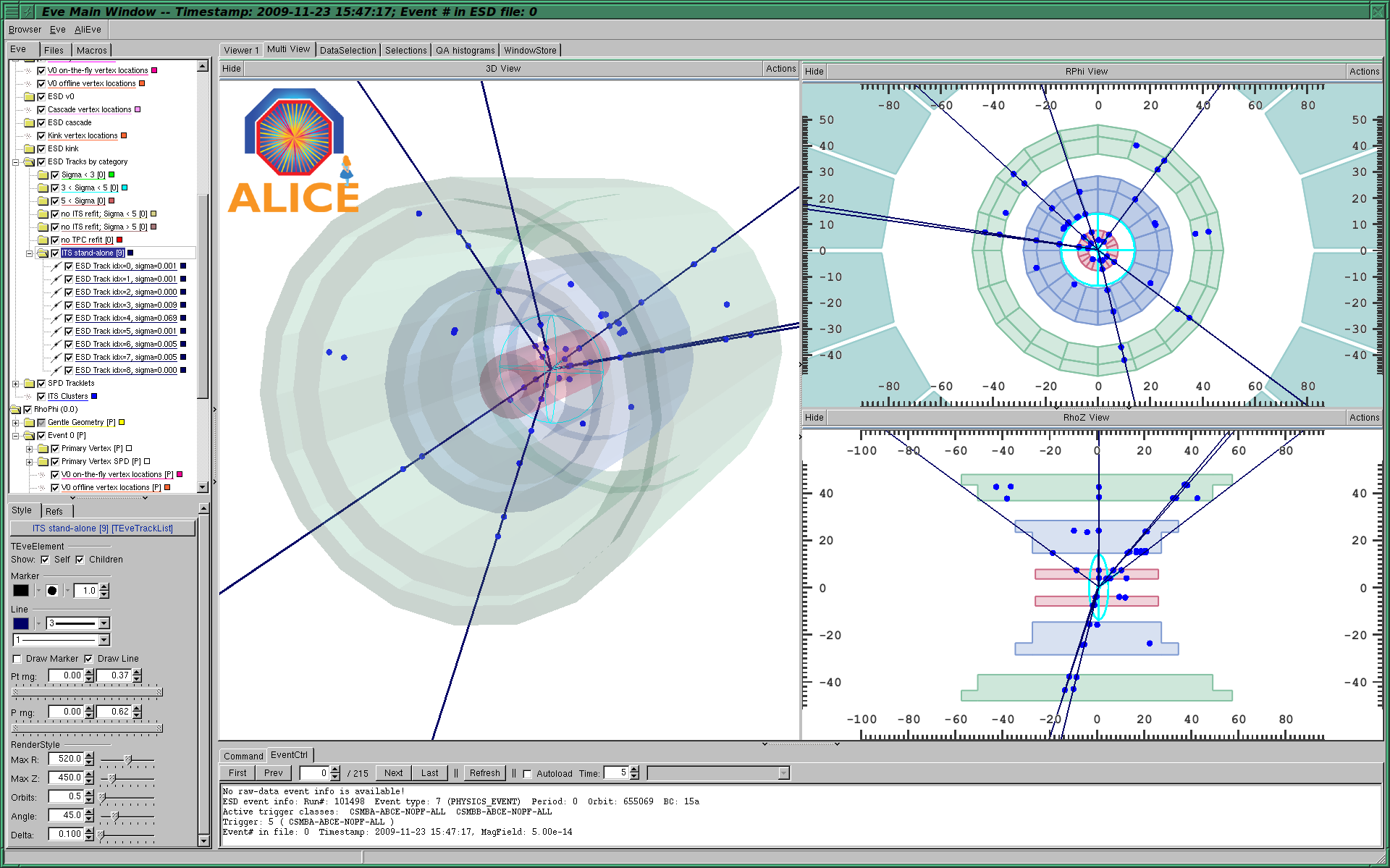}
\caption{The first pp collision candidate shown by the event display in the
  ALICE counting room (3D view, $r$-$\phi$ and $r$-$z$ projections), the
  dimensions are shown in cm.
  The dots correspond to hits in the silicon vertex detectors (SPD, SDD and SSD), the lines correspond to tracks reconstructed using loose quality cuts. The ellipse drawn in the middle of the detector surrounds the reconstructed event vertex.}
\label{figEvent}
\vspace{0.5cm}
\includegraphics[width=0.95\textwidth]{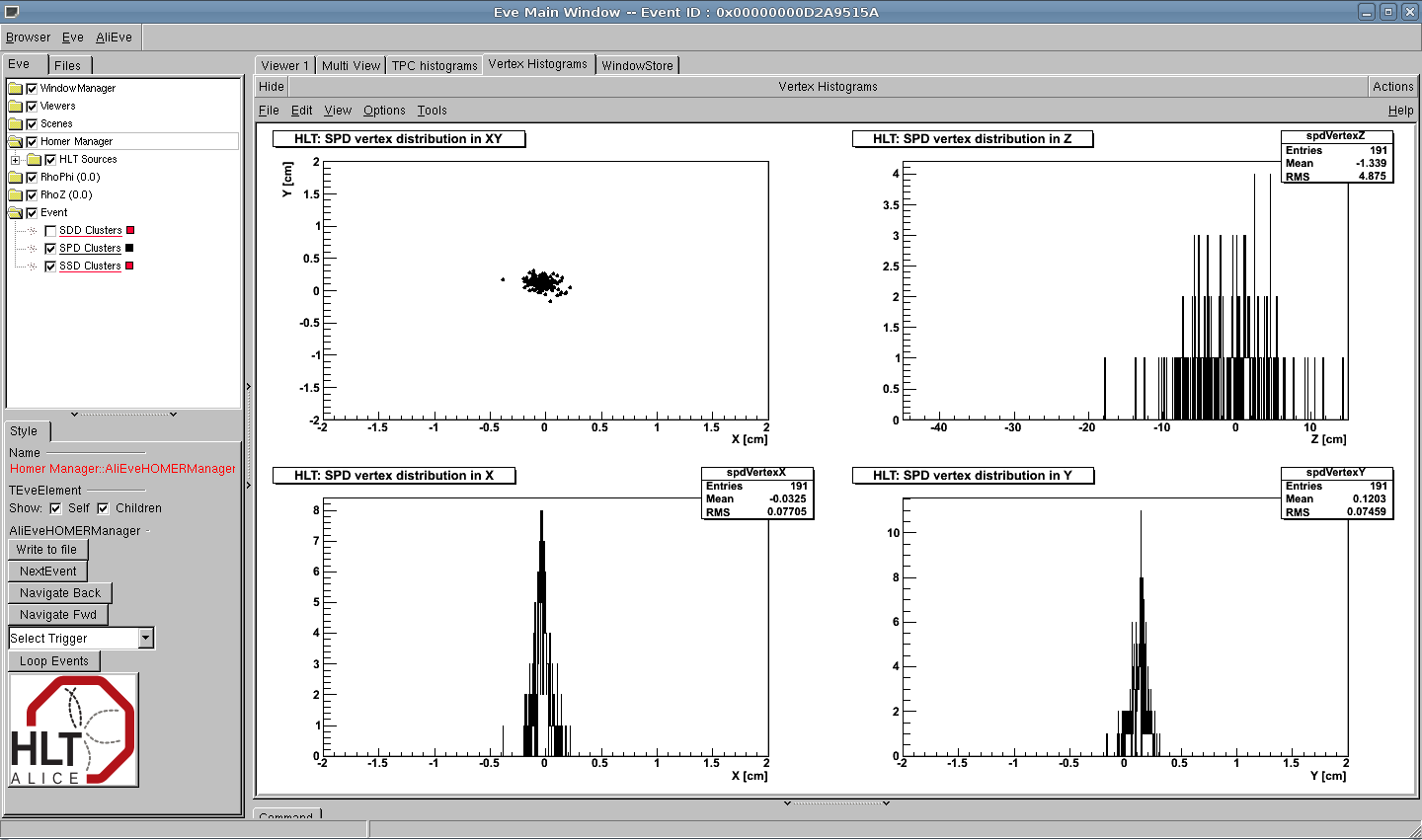}
\caption{Online display of the vertex positions reconstructed by the High-Level
  Trigger (HLT). The figure shows, counter-clockwise from top left, the
  position in the transverse plane for all events with a reconstructed vertex, the projections along the transverse coordinates $x$ and $y$, and the distribution along the beam line ($z$-axis).}
\label{figHLT}
\end{figure*}

\section{LHC and the run conditions}

The LHC, built at CERN in the
circular tunnel of 27~km circumference previously used by the Large Electron--Posi-tron collider (LEP), will provide the highest energy ever explored with particle accelerators. It is designed to collide two counter-rotating beams of protons or heavy ions.
The nominal
centre-of-mass energy for proton--proton collisions is $\unit[14]{TeV}$. However, collisions can be obtained down to $\sqrt{s} = \unit[900]{GeV}$, which corresponds to the beam injection energy.

The results from the first proton--proton collisions presented here were obtained  during the early commissioning phase of the LHC, when two proton bunches were circulating for the first time concurrently in the machine.
The bunches used were the so-called ``pilot bunches'': low intensity bunches used during machine commissioning, with a few $10^{9}$ protons per bunch.
The two beams were brought into nominal position for collisions without a specific attempt to maximize the interaction rate.
The nominal r.m.s. size of LHC beams at injection energy is about $\unit[300]{\mu m}$ in the transverse direction and $\unit[10.5]{cm}$ in the longitudinal ($z$-axis) direction. However, at this early stage, the beam parameters can deviate from these nominal values; they were not measured for the fill used in this analysis. For the previous fill, for which the longitudinal size was measured, it was found to be shorter, with an r.m.s. of about 8 cm. Assuming Gaussian beam profiles, the luminous region should be smaller than the beam size by a factor of $\sqrt{2}$ in all directions.

Shortly after circulating beams were established, the ALICE data aquisition
system~\cite{DAQ} started collecting ev\-ents with a trigger based on the
Silicon Pixel Detector (SPD), requiring two or more hits in the SPD in
coincidence with the passage of the two colliding bunches as inferred from
beam pickup detectors. As a precaution, only a small subset of the detector subsystems, including the silicon tracking detectors and the scintillator trigger counters, was turned on, in order to assess the beam conditions provided by the LHC.

The trigger rate was measured just before collisions with the same trigger
conditions. Without beams we measured a rate of $3 \times 10^{-4}$\,Hz (in coincidence with one bunch crossing interval per orbit). In coincidence with the passage of the bunch of one circulating beam the rate was  $0.006$\,Hz. As soon as the second beam was injected in the accelerator, the event rate increased significantly, to $0.11$\,Hz.
The first event that was analyzed and displayed in the counting room by
the offline reconstruction software AliRoot~\cite{AliRoot} running in online mode is shown in
Fig.~\ref{figEvent}. This marked symbolically  the keenly anticipated start
of the physics exploitation of the ALICE
experiment\footnote{The event display started
  shortly after data taking and therefore missed the first few
  events.}. The online reconstruction software implemented in the
High-Le\-vel Trigger (HLT) computer farm~\cite{HLT} also analyzed the events
in real time and calculated the vertex position of the collected events,
shown in Fig.~\ref{figHLT}. The distributions are very narrow in the
transverse plane (sub-millimetre,
including contributions from detector resolution and residual misalignment), of about the expected size in the longitudinal direction and  well positioned with respect to the nominal centre of the ALICE detector.
This provided immediate evidence that a substantial fraction of the events corresponded to collisions between the protons of the two counter-rotating beams.

After 43 minutes, the two beams were dumped in order to proceed with the LHC commissioning programme. In total, 284 events were triggered and recorded during this short, but important, first run of the ALICE experiment with colliding beams.

\section{The ALICE experiment}

ALICE, designed as the dedicated heavy-ion experiment at the LHC, also has excellent performance for
proton--proton interactions~\cite{ALICE_PPR}. The experiment consists of a large number
of detector subsystems~\cite{ALICEdet} inside a solenoidal magnet ($B = 0.5$~T). The magnet was off during this run.

During the several months of running with cosmic rays in 2008 and 2009, all of the ALICE detector subsystems were extensively commissioned, calibrated and used for data taking~\cite{commiss}. Data were collected for an initial alignment of the parts of the detector that had sufficient exposure to the mostly vertical cosmic ray flux. Data were also taken during various LHC injection tests to perform timing measurements and other calibrations.

\begin{figure*}[t!]
\centering
\includegraphics[width=0.49\linewidth]{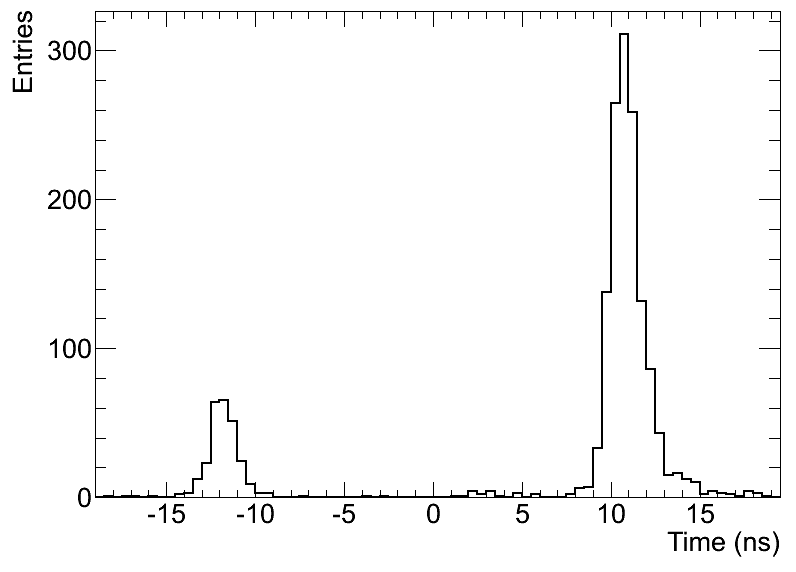}
\includegraphics[width=0.49\linewidth]{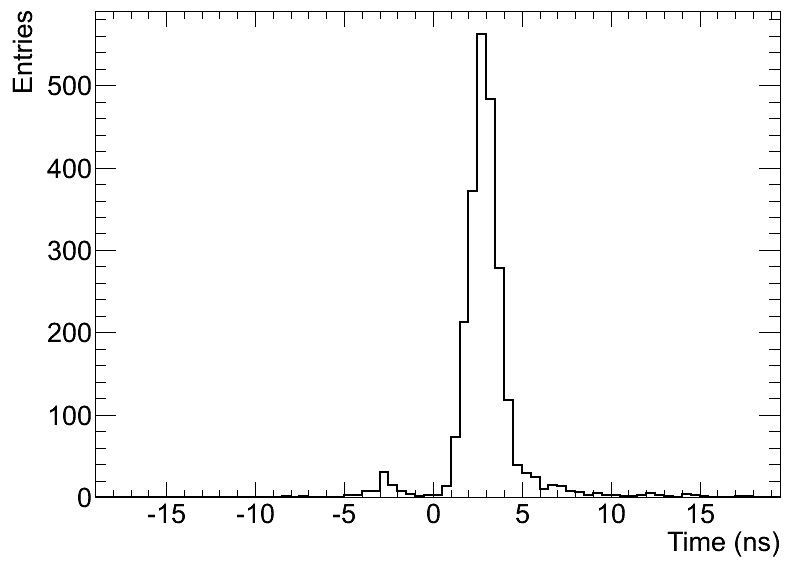}
\caption{Arrival time of particles in the VZERO detectors relative to the
  beam crossing time (time zero). A number of beam-halo or beam--gas events
  are visible as secondary peaks in VZERO-A (left panel) and VZERO-C (right
  panel). This is because particles produced in background interactions
  arrive at earlier times in one or the other of the two counters.
The majority of the signals have the correct arrival time expected for collisions around the nominal vertex.}
\label{figV0}
\end{figure*}

Collisions take place at the centre of the ALICE detector,
inside a beryllium vacuum beam pipe ($\unit[3]{cm}$ in radius
and $\unit[800]{\mu m}$ thick).
The tracking system in the ALICE central barrel  covers the full azimuthal range in the pseudorapidity
window $|\eta| < 0.9$.
 It has been designed to cope with the highest charged-particle densities
 expected in central Pb--Pb collisions. The following four detector
 subsystems were active during data taking and were used in this analysis.
\begin{itemize}
\item The Silicon Pixel Detector (SPD) consists of two cylindrical
layers with radii of $3.9$ and $\unit[7.6]{cm}$ and has
about $9.8$ million pixels of size $50 \times 425 \, \mu$m$^2$. It covers the
pseudorapidity ranges $|\eta|<2$ and $|\eta|<1.4$ for the inner and outer layers respectively, for particles originating at the centre of the detector. The effective $\eta$-acceptance is larger due to the longitudinal spread of the position of the interaction vertex. The detector is read out by custom-designed ASICs bump-bonded directly on silicon ladders. Each chip contains 8192 channels and also provides a fast trigger signal if at least one of its pixels is hit. The trigger signals from all 1200 chips are then combined in a programmable logic unit which provides a level-0 trigger signal to the central trigger processor.
The total thickness of the SPD amounts to about 2.3\,\% of a radiation length. About 83\,\% of the channels were operational for particle detection and 77\,\% of the chips were used in the trigger logic.
The SPD was aligned using cosmic-ray tracks collected during 2008~\cite{AlignmentNote}, and the residual misalignment was estimated to be below 10~$\mu$m for the modules well covered by mostly vertical tracks. The modules on the sides are likely to be affected by larger residual misalignment.
\item The Silicon Drift Detector (SDD) consists of two cylindrical layers at radii of $15.0$ and $\unit[23.9]{cm}$ and covers the region $|\eta| < 0.9$. It is composed of 260 sensors with an internal voltage divider providing a drift field of 500~V/cm and MOS charge injectors that allow measurement of the drift speed via dedicated calibration triggers. The charge signal of each of the 133\,000 collection anodes, arranged with a pitch of $\unit[294]{\mu \mathrm m}$, is sampled every 50~ns by an ADC in the front-end electronics. The total thickness of the SDD layers (including mechanical supports and front-end electronics) amounts to 2.4\,\% of a radiation length.  About 92\,\% of the anodes were fully operational.
\item The two layers of the double-sided Silicon Strip Detector (SSD) are located at radii of $38$ and $\unit[43]{cm}$ respectively, covering $|\eta|<0.97$. The SSD consists of 1698 sensors with a strip pitch of $\unit[95]{\mu \mathrm m}$ and a stereo angle of 35~mrad. The detector provides a measurement of the charge deposited in each of its $2.5 \times 10^6$ strips. The position resolution is better than $\unit[20]{\mu \mathrm m}$ in the $r$-$\varphi$ direction and about 0.8~mm in the direction along the beam line. The thickness of the SSD, including supports and services, corresponds to 2.2\,\% of a radiation length. About 90\,\% of the SSD area was active during data taking.
\item The VZERO detector consists of two arrays of 32
scintillators each, which are placed around the beam pipe on either side of
the interaction region: VZERO-A at $z = 3.3$~m, covering the
pseudorapidity range $2.8 < \eta < 5.1$, and VZERO-C at $z =
-0.9$~m, covering the pseudorapidity range $-3.7 < \eta < -1.7$. The time
resolution of this detector is better than $\unit[1]{ns}$.
Its response is recorded in a time window of $\pm$25~nsec around the
nominal beam crossing time. For events collected in this run, the
arrival times of particles at the detector relative to this ``time zero''
is shown in Fig.~\ref{figV0}.
Note that in
general several particles are registered for each event.
Particles hitting one of the detectors  before the
beam crossing have negative arrival times and are typically due to
interactions taking place outside the central region of ALICE.

\end{itemize}
More details about the ALICE experiment and its detector subsystems can be found in~\cite{ALICEdet}.

The trigger used to record the events for the present analysis is defined by requiring at least two hit chips in the SPD, in coincidence with the signals from the two beam pick-up counters indicating the presence of two passing proton bunches.
The efficiency of this trigger as well as all other corrections have been
studied using two different Monte Carlo generators, PYTHIA 6.4.14~\cite{PYTHIA} tune D6T~\cite{D6Ttune} and PHOJET~\cite{PHOJET}, for INEL and NSD interactions. The trigger efficiencies for non-diffractive, single-diffractive, and double-diffractive events were evaluated separately, and found to be 98--99\,\%, 48--58\,\%, and 53--76\,\% respectively. The ranges are determined by the two event generators. These event classes were combined for the corrections using the fractions measured by UA5~\cite{UA5diff}: non-diffractive $0.767 \pm 0.059$; single-diffractive $0.153 \pm$ \linebreak $\pm 0.031$; double-diffractive $0.08 \pm 0.05$.
The resulting efficiencies were found to be 87--91\,\% for the INEL normalization and 94--97\,\% for the NSD normalization, again depending on the event generator used.

The results presented in the following sections are those obtained with
PYTHIA. The difference between results corrected with PY\-THIA and PHOJET is used in the estimate
of the systematic uncertainty.

\section{Data analysis}

\label{analysis}

The data sample used in the present analysis consists of 284
events recorded without magnetic field.
The results presented here are based on the analysis of the SPD data. However, information from the SDD, SSD and VZERO was used to crosscheck the identification and removal of background events.

\begin{figure}[t!]
\centering
\includegraphics[width=\linewidth]{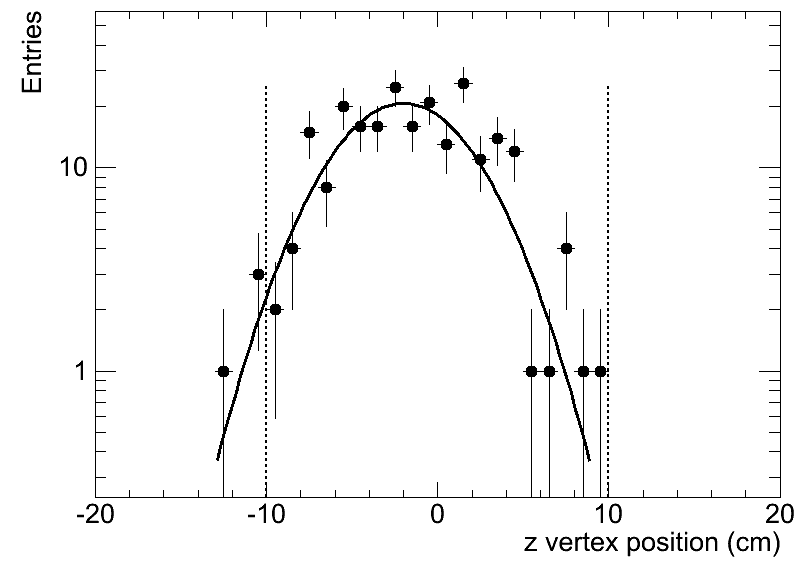}
\caption{Longitudinal vertex distribution from hit correlations in the
two pixel layers of the ALICE inner tracking system.
Vertical dashed lines indicate the region $|z|<10$~cm, where the events for
the present analysis are selected. A Gaussian fit with an estimated
r.m.s. of about $\unit[4]{cm}$ to the central part is also shown.}
\label{figVtx}
\end{figure}

In the SPD analysis, the position of the interaction vertex is
reconstructed~\cite{VertexNote} by correlating hits in the two silicon-pixel layers to obtain
tracklets. The achieved resolution depends on the track multiplicity and for this specific vertex reconstruction is approximately $0.1$--$\unit[0.3]{mm}$ in
the longitudinal direction and $0.2$--$\unit[0.5]{mm}$ in the transverse direction. For events with only one charged track, the vertex position is determined by intersecting the SPD tracklet with the mean beam axis determined from the vertex positions of other events in the sample.
A vertex was reconstructed in 94\,\% of the selected
events. The distribution of the vertex position in the longitudinal direction ($z$-axis) is shown in Fig.~\ref{figVtx}.
For events originating from the centre of the detector, the vertex-reconstruction efficiency was estimated, using Monte Carlo simulations, to be 84\,\% for INEL interactions and
92\,\% for NSD collisions. These efficiencies decrease for larger $|z|$-values of the vertex in low-multiplicity events; therefore, only events with vertices within $|z|<\unit[10]{cm}$ were used.
This allows for an accurate charged-particle density measurement in the pseudorapidity range $|\eta| < 1.6$ using both SPD layers.

Using the reconstructed vertex as the origin, we calculate the differences in azimuthal ($\Delta \varphi$, bending plane) and polar ($\Delta \theta$, non-bending direction) angles of pairs of hits with one hit in each SPD layer.
These tracklets~\cite{TrackletNote} are selected by a cut on the sum of the squares of $\Delta \varphi$ and $\Delta \theta$, each normalized to its estimated resolution (\unit[80]{mrad} and \unit[25]{mrad}, respectively).
When more than one hit in a layer matches a hit
in the other layer, only the hit combination with the smallest angular
difference is used. This occurs in only 2\,\% of the matched
hits.

The number of primary charged particles is estimated by counting
the number of tracklets. This number was corrected for:
\begin{itemize}
\item{trigger inefficiency;}
\item{detector and reconstruction inefficiencies;}
\item{contamination by decay products of long-lived particles (K$^0_{\rm s}$, $\Lambda$, etc.), gamma conversions and  secondary interactions.}
\end{itemize}
The corrections are determined as a function of the $z$-position of the
primary vertex, and on the pseudorapidity of the tracklet. For the analyzed sample the average correction factor for tracklets is about 1.5.

The beam--gas and beam-halo background events were removed by a cut on the
ratio between the number of tracklets and the total number of hits in the
tracking system (SPD, SDD, and SSD); this ratio is smaller for background events
(as measured in the previous fills triggering on the bunch passage from one
side) than for collisions~\cite{BckNote}. In addition, the timing information from the
VZERO detector was used for background rejection by removing events with
negative arrival time (see Fig.~\ref{figV0}). The event quality and event classification was crosschecked by a visual scan of the whole event sample.
In total 29 events (i.e. about 10\,\%) were rejected as beam induced background, which is consistent with the rate expected from previous fills. The remaining background was estimated from the vertex distribution and found to be negligible.
 The contamination from coincidence with a cosmic event was estimated to be
 one event in the full sample. Indeed, two cosmic events were identified by scanning, both without reconstructed vertex.

\begin{figure}[t!]
\centering
\includegraphics[width=\linewidth]{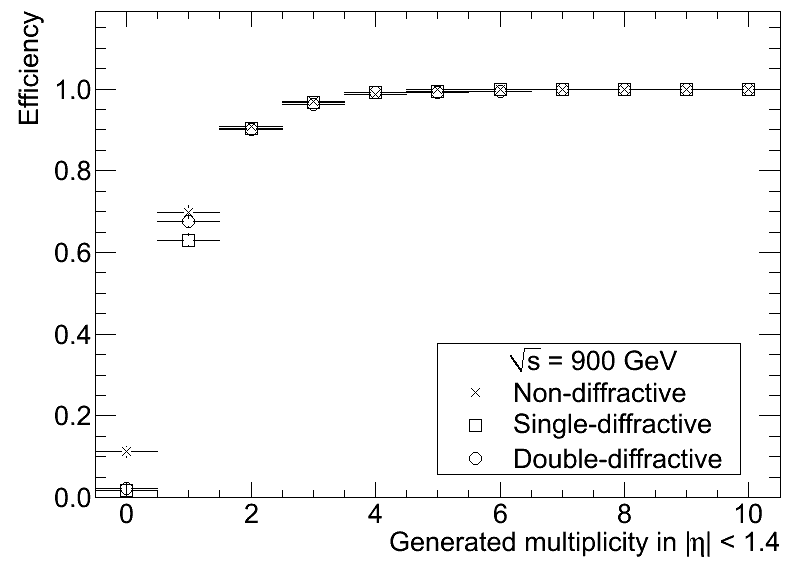}
\caption{Multiplicity dependence of the combined efficiency to select an event as minimum bias and to reconstruct its vertex in SPD, for non-diffractive (crosses), single-diffractive (squares), and double-diffractive (circles) events, based on PYTHIA events.}
\label{figverteff}
\end{figure}

Particular attention has been paid to events having zero or one charged tracklets in the SPD acceptance. The vertex-finding efficiency for events with one charged particle in the acceptance is about 80\,\%. The number of zero-track events has been estimated by Monte Carlo calculations.
The total number of collisions used for the
normalization was calculated from the number of events selected for the
analysis, corrected for the vertex-reconstruction inefficiency.
In order to obtain the normalization for INEL and NSD events, we further corrected the number of selected events for the trigger efficiency for these two event classes. In addition, for NSD events, we subtract the single-diffractive contribution. These corrections, as well as those for the vertex finding efficiency, depend on the event char\-ged-particle multiplicity, see
Fig.~\ref{figverteff}.
The dependence of the event-finding efficiency (combining event selection
and vertex finding) on multiplicity was calculated for
different interaction types using our detector simulation,
and is above 98\,\% for
events with at least two charged particles.
The averaged combined corrections for the vertex reconstruction
efficiency and the selection efficiency is 20\,\% for INEL interactions and much smaller for NSD interactions, due to the cancelation of some contributions.

The various corrections mentioned above
  were calculated using the full GEANT 3~\cite{GEANT} simulation of the ALICE detector as
  included in the offline framework AliRoot. In order to estimate the systematic uncertainties, the above analysis was repeated by:
\begin{itemize}
\item{applying different cuts for the tracklet definition (varying the angle cut-off by $\pm 50$\,\%);}
\item{varying by $\pm 10$\,\% the density of the material in the tracking system, thus changing the material budget;}
\item{using the non-aligned geometry;}
\item{varying by $\pm 30$\,\% the composition of the produced particle types with respect to the yields suggested by the event generators;}
\item{varying the particle yield below $\unit[100]{MeV/c}$ by $\pm 30$\,\%;}
\item{evaluating the uncertainty in the normalization to \linebreak
 INEL and NSD
  samples by varying the ratios of the non-diffractive, single-diffractive
  and double-diffractive \linebreak cross sections according to their measured values and errors~\cite{UA5diff}
    and using two different models for diffraction kinematics (PYTHIA and PHOJET).}
\end{itemize}

An additional source of systematic error comes from the limited statistics used so far to determine the efficiencies of the SPD detector modules. In test beams, the SPD efficiency in active areas was measured to be higher than $99.8\,\%$. This was crosschecked in-situ with cosmic data, but only over a limited area and with limited statistics. At this stage, we have assigned a conservative value of 4\,\% to this uncertainty.
The triggering efficiency of the SPD was estimated from the data itself,
using the trigger information recorded in the data stream for events with
more than one tracklet, and found to be very close to 100\,\%, with an error of about 2\,\% (due to the limited statistics).

These contributions to the systematic uncertainty on the charged particle pseudorapidity density are summarized in Table~\ref{systable}.
Our conclusion is that the total systematic
uncertainty on the pseudorapidity density is less than $\pm 7.2$\,\% for INEL collisions and $\pm 7.1$\,\% for NSD collisions. The largest contribution comes from uncertainties in cross sections of diffractive processes and their kinematic simulation.

\begin{table}[t!]
\centering
\caption{Contributions to systematic uncertainties on the measurement of the charged-particle pseudorapidity density.
}
\label{systable}
\begin{tabular}{lc}
	\hline
    \tblspc
	Uncertainty					&  \\
	\hline
    \tblspc
    Tracklet selection cuts  	& negl.  \\
    Material budget 			& negl.  \\
    Misalignment          & 0.5\,\% \\
    Particle composition            	& negl.	\\
    Transverse-momentum spectrum      	& 0.5\,\%	\\
	Contribution of diffraction (INEL)  	& 4\,\%	\\
	Contribution of diffraction (NSD)   	& 4.5\,\%	\\
	Event-generator dependence (INEL)	& 4\,\%	\\
	Event-generator dependence (NSD)	& 3\,\% 	\\
    Detector efficiency & 4\,\%	 \\
    SPD triggering efficiency & 2\,\% \\
	Background events    		& negl.    \\
	\hline
    \tblspc
	Total (INEL)			& 7.2\,\%  \\
	Total (NSD) 			& 7.1\,\%  \\
	\hline
\end{tabular}
\end{table}

More details about this analysis, corrections, and the evaluation of the systematic uncertainties can be found in~\cite{JanFiete}.

\vspace{0.8cm}

\section{Results}

\begin{table*}[t!]
\centering
\caption{Comparison of charged primary particle pseudorapidity densities at central pseudorapidity ($|\eta|<0.5$) for inelastic (INEL) and non-single diffractive (NSD) collisions measured by the ALICE detector in ${\rm p}{\rm p}$ interactions and by UA5 in ${\rm p} \overline{\rm p}$ interactions~\cite{UA5_dNdEta} at a centre-of-mass energy of $\unit[900]{GeV}$. For ALICE, the first error is statistical and the second is systematic; no systematic error is quoted by UA5. The experimental data are also compared to the predictions for ${\rm p}{\rm p}$ collisions from different models. For PYTHIA the tune versions are given in parentheses. The correspondence is as follows: D6T is tune (109); ATLAS CSC is tune (306); Perugia-0 is tune (320).}
\label{tabdNdeta}
\begin{tabular}{lccccccc}
  \hline\noalign{\smallskip}
  Experiment & ALICE pp & UA5 p$\overline{\rm p}$~\cite{UA5_dNdEta}  & QGSM~\cite{QGSMcal} & \multicolumn{3}{c}{PYTHIA~\cite{PYTHIA}} & PHOJET~\cite{PHOJET} \\
  Model   &  &   &  &  (109)~\cite{D6Ttune} &   (306)~\cite{CSCtune}   &   (320)~\cite{Perugiatune} &  \\
  \noalign{\smallskip}\hline\noalign{\smallskip}
  INEL &   $3.10 \pm 0.13 \pm 0.22$ & $3.09 \pm 0.05$ &2.98&2.33&2.99&2.46&3.14\\
  NSD  &   $3.51 \pm 0.15 \pm 0.25$ & $3.43 \pm 0.05$ &3.47&2.83&3.68&3.02&3.61\\
  \noalign{\smallskip}\hline
\end{tabular}
\end{table*}

\begin{figure}[t!]
\centering
\includegraphics[width=\linewidth]{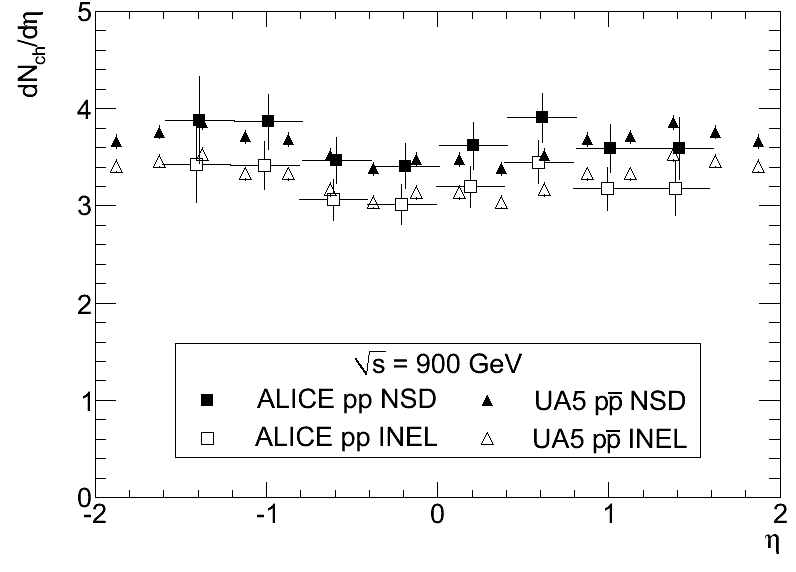}
\caption{Pseudorapidity dependence of \dNdEta for INEL and NSD
collisions. The ALICE measurements (squares) are compared to UA5 data (triangles)~\cite{UA5_dNdEta}. The errors shown are statistical only.}
\label{figdNdEta}
\end{figure}

Figure~\ref{figdNdEta} shows the charged primary particle pseudorapidity density distributions obtained for INEL and NSD interactions in the range $|\eta| < 1.6$. The pseudorapidity density obtained in the central region $|\eta| < 0.5$ for INEL interactions is $3.10 \pm 0.13(\emph{stat.}) \pm 0.22(\emph{syst.})$ and for NSD interactions is $3.51 \pm 0.15(\emph{stat.}) \pm 0.25(\emph{syst.})$.
Also shown in Fig.~\ref{figdNdEta} are the previous measurements of proton--antiproton interactions from the UA5 experiment~\cite{UA5_dNdEta}. Our results obtained for proton--proton
interactions are consistent with those for proton--antiproton
interactions, as expected from the fact that the predicted difference
(0.1--0.2\,\%) is well below measurement uncertainties. The measurements at central pseudorapidity ($|\eta|<0.5$) are
summarized in Table~\ref{tabdNdeta} together with
 model predictions obtained with QGSM, PHOJET and three different PYTHIA tunes.
 PYTHIA 6.4.14, tune D6T, and PHOJET yield respectively the lowest and
 highest charged particle densities.
 Therefore, these two have been used for the evaluation of our systematic
 errors.
 PYTHIA 6.4.20, tunes ATLAS CSC and Perugia-0, are candidates for use
 by the LHC experiments at higher LHC energies and are shown for
 comparison.

\begin{figure}[t!]
\centering \includegraphics[width=\linewidth]{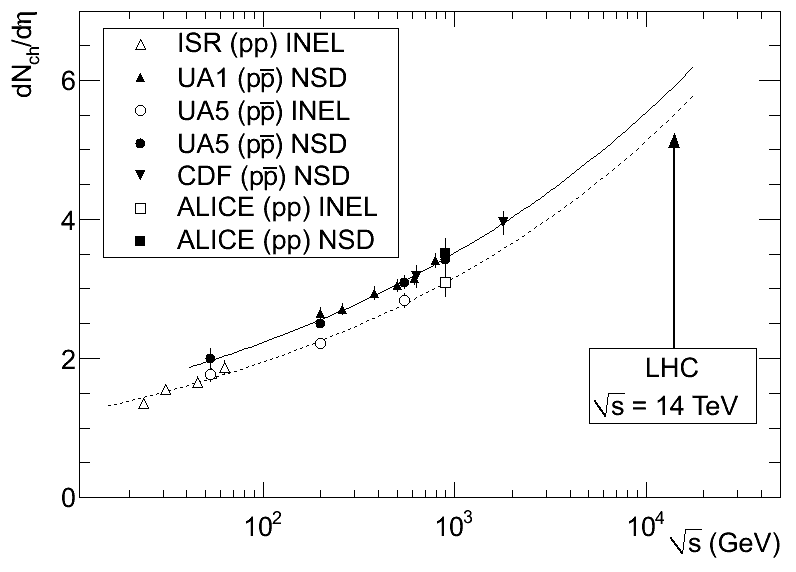}
\caption{Charged-particle pseudorapidity density in the central rapidity region
in proton--proton and proton--antiproton interactions as a function of
the centre-of-mass energy. The dashed and solid lines (for INEL and NSD interactions respectively) indicate the
fit using a power-law dependence on energy.}
\label{figdNdEtaSqrtS}
\end{figure}

Figure \ref{figdNdEtaSqrtS} shows the centre-of-mass energy dependence of the
pseudorapidity density in the central region ($|\eta|<0.5$).
The data points are obtained in the $|\eta|<0.5$ range from this experiment and from references~\cite{UA5_dNdEta,R210,ITS_dNdEta,UA1,UA5Rep,CDF_dNdEta}, and are corrected for differences in pseudorapidity range where necessary, fitting the pseudorapidity distribution around $\eta = 0$. As noted above, there is good agreement between pp and p$\overline{\mathrm p}$ data at the same energy. The dashed and solid lines (for INEL and NSD interactions respectively) are obtained by fitting the density of charged particles in the central pseudorapidity rapidity region with a power-law dependence on energy.

Using this parametrization, the extrapolation to the
nominal LHC energy of $\sqrt{s}=\unit[14]{TeV}$ yields
\dndeta{5.5} and \dndeta{5.9} for INEL
and NSD interactions respectively.

\section{Conclusion}

Proton--proton collisions observed with the ALICE detector in the early phase of the LHC commissioning have been used to measure the pseudorapidity density of charged primary particles at $\sqrt{s} = 900$~GeV. In the central pseudorapidity region ($|\eta|<0.5$), we obtain \dndeta{3.10 \pm 0.13 {\mathrm(stat.)} \pm 0.22 {\mathrm(syst.)}} for all
inelastic and
\dndeta{3.51 \pm 0.15 {\mathrm(stat.)} \pm 0.25 {\mathrm(syst.)}} for non-single diffractive proton--proton interactions.
The results are consistent with earlier measurements of primary
charged-particle production in proton--antiproton interactions at the same energy. They are also compared with model calculations.

These results have been obtained with a small sample of events during
the early commissioning of the LHC. They demonstrate that the LHC and its
experiments have finally entered the phase of physics exploitation, within
days of starting up the accelerator complex in November 2009.

\begin{acknowledgement}

\section*{Acknowledgements}

The ALICE collaboration would like to thank all its engineers and technicians for their invaluable contributions to the construction of the experiment. We would like to thank and congratulate the CERN accelerator teams for the outstanding performance of the LHC complex at start up, and for providing us with the collisions used for this paper on such a short notice!

The ALICE collaboration acknowledges the following funding agencies for their support in building and
running the ALICE detector:
\begin{itemize}
\item{}
Calouste Gulbenkian Foundation from Lisbon and Swiss Fonds Kidagan, Armenia;
\item{}
Conselho Nacional de Desenvolvimento Cientfico e Tecnolgico (CNPq), Financiadora de Estudos e Projeto (FINEP),
Funda\c{c}\~{a}o de Amparo \`{a} Pesquisa do Estado de S\~{a}o Paulo (FAPESP);
\item{}
National Natural Science Foundation of China (NSFC), the Chinese Ministry of Education (CMOE)
and the Ministry of Science and Technology of China (MSTC);
\item{}
Ministry of Education and Youth of the Czech Rebublic;
\item{}
Danish National Science Research Council and the Carlsberg Foundation;
\item{}
The European Research Council under the European Community's Seventh Framework Programme;
\item{}
Helsinki Institute of Physics and the Academy of Finland;
\item{}
French CNRS-IN2P3, the `Region Pays de Loire', `Region Alsace', `Region Auvergne' and CEA, France;
\item{}
German BMBF and the Helmholtz Association;
\item{}
Hungarian OTKA and National Office for Research and Technology (NKTH);
\item{}
Department of Atomic Energy and Department of Science and Technology of the Government of India;
\item{}
Istituto Nazionale di Fisica Nucleare (INFN) of Italy;
\item{}
MEXT Grant-in-Aid for Specially Promoted Research, Ja\-pan;
\item{}
Joint Institute for Nuclear Research, Dubna;
\item{}
Korea Foundation for International Cooperation of Science and Technology (KICOS);
\item{}
CONACYT, DGAPA, M\'{e}xico, ALFA-EC and the HELEN Program (High-Energy physics Latin-American--European Network);
\item{}
Stichting voor Fundamenteel Onderzoek der Materie \linebreak (FOM) and the Nederlandse Organistie voor Wetenschappelijk Onderzoek (NWO), Netherlands;
\item{}
Research Council of Norway (NFR);
\item{}
Polish Ministry of Science and Higher Education;
\item{}
National Authority for Scientific Research - NASR (Autontatea Nationala pentru Cercetare Stiintifica - ANCS);
\item{}
Federal Agency of Science of the Ministry of Education and Science of Russian Federation, International Science and
Technology Center, Russian Federal Agency of Atomic Energy, Russian Federal Agency for Science and Innovations and CERN-INTAS;
\item{}
Ministry of Education of Slovakia;
\item{}
CIEMAT, EELA, Ministerio de Educaci\'{o}n y Ciencia of Spain, Xunta de Galicia (Conseller\'{\i}a de Educaci\'{o}n),
CEA\-DEN, Cubaenerg\'{\i}a, Cuba, and IAEA (International Atomic Energy Agency);
\item{}
Swedish Reseach Council (VR) and Knut $\&$ Alice Wallenberg Foundation (KAW);
\item{}
Ukraine Ministry of Education and Science;
\item{}
United Kingdom Science and Technology Facilities Council (STFC);
\item{}
The United States Department of Energy, the United States National
Science Foundation, the State of Texas, and the State of Ohio.
\end{itemize}
\end{acknowledgement}

%

\begin{thebibliography}{00}

\bibitem{LHC} L.~Evans and P.~Bryant (editors), JINST \textbf{3} (2008) S08001



\bibitem{ALICEdet} ALICE Collaboration, K.~Aamodt et al., JINST \textbf{3} (2008) S08002


\bibitem{UA5_dNdEta} UA5 Collaboration,
   G.J.~Alner et al., Z.~Phys.~C \textbf{33} (1986) 1

\bibitem{QGSM} A.B.~Kaidalov, Phys.~Lett.~B \textbf{116} (1982) 459;
A.B.~Kaidalov and K.A.~Ter-Martirosyan, Phys.~Lett.~B \textbf{117} (1982) 247;
A.B.~Kaidalov and K.A.~Ter-Martirosyan, Yad.~Fiz. \textbf{39} (1984) 1545 and Sov.~J.~Nucl.~Phys. \textbf{39} (1984) 979;
    A.B.~Kaidalov and K.A.~Ter-Martirosyan, Yad.~Fiz. \textbf{40} (1984) 211 and Sov.~J.~Nucl.~Phys. \textbf{40} (1984) 135

\bibitem{DPM} A.~Capella et al., Z.~Phys.~C \textbf{3} (1980) 329;
A.~Capella and J.~Tran~Thanh~Van, Z.~Phys.~C \textbf{10} (1981) 249;
A.~Capella and J.~Tran~Thanh~Van, Phys.~Lett.~B \textbf{114} (1982) 450;
A.~Capella, U.~Sukhatme, C.-I.~Tan and J.~Tran~Thanh~Van, Phys.~Rep. \textbf{236} (1994) 225




\bibitem{QGSMMC} N.S.~Amelin and L.V.~Bravina, Yad.~Fiz. \textbf{51} (1990) 211 and Sov. J. Nucl. Phys. \textbf{51} (1990) 133;
    N.S.~Amelin et al., Yad.~Fiz. \textbf{51} (1990) 512 and Sov.~J.~Nucl.~Phys. \textbf{51} (1990) 327;
    N.S.~Amelin et al., Yad.~Fiz. \textbf{52} (1990) 272 and Sov.~J.~Nucl.~Phys. \textbf{51} (1990) 172

\bibitem{DPMJet} P.~Aurenche et al., Phys.~Rev.~D \textbf{45} (1992) 92



\bibitem{PHOJET} R.~Engel, J.~Ranft and S.~Roesler, Phys.~Rev.~D \textbf{52} (1995) 1459

\bibitem{difpap} J.G.~Rushbrooke and B.R.~Webber, Phys.~Rep.~C \textbf{44} (1978) 1

\bibitem{R210} UA5 Collaboration, K.~Alpg{\aa}rd et al., Phys.~Lett.~B \textbf{112} (1982) 183;
M.~Ambrosio et al., AIP Conference Proceedings \textbf{85} (1982) 602

\bibitem{DAQ} T.~Anti\v{c}i\'{c} et al., ALICE Internal Note ALICE-INT-2005-015 (2005); T.~Anti\v{c}i\'{c} et al. (ALICE Collaboration), J. Phys. Conf. Series \textbf{119} (2008) 022006

\bibitem{AliRoot} ALICE Collaboration, AliRoot, ALICE Offline simulation, reconstruction and analysis framework, http://aliceinfo.cern.ch/Offline/

\bibitem{HLT} T.M.~Steinbeck (ALICE Collaboration), Proceedings of CHEP 2009, March 2009, Prague, to be published in J. Phys. Conf. Series


\bibitem{ALICE_PPR} ALICE Collaboration,
  J.~Phys.~G \textbf{30} (2004) 1517 and
  J.~Phys.~G \textbf{32} (2006) 1295

\bibitem{commiss} P.G.~Kuijer (ALICE Collaboration),
  Nucl. Phys. A \textbf{830} (2009) 81C; F.~Prino (ALICE Collaboration),
  Nucl. Phys. A \textbf{830} (2009) 527C; R.~Santoro et al. (ALICE Collaboration), JINST \textbf{4} (2009) P03023; G.~Aglieri Rinella et al,
  CERN preprint CERN-2008-008 (2008); P.~Giubellino (ALICE Collaboration), Proceedings of EPS HEP 2009, July 2009 Krakow, to be published in Proc. of Science






\bibitem{AlignmentNote}
C.~Bombonati et al., ALICE Internal Note ALICE-INT-2009-035 (2009)

\bibitem{PYTHIA}
  T. Sj\"{o}strand, Comput.~Phys.~Commun. \textbf{82} (1994) 74;
  T. Sj\"{o}strand, S.~Mrenna and P.~Skands, JHEP \textbf{2006} 05 (2006) 026

\bibitem{D6Ttune} M.G.~Albrow et al. (Tev4LHC QCD Working Group), arXiv:hep-ph/0610012 (2006), D6T (109) tune

\bibitem{UA5diff} UA5 Collaboration, R.E.~Ansorge et al., Z. Phys. C \textbf{33} (1986) 175

\bibitem{VertexNote} E.~Bruna et al., ALICE Internal Note ALICE-INT-2009-018 (2009)

\bibitem{TrackletNote} D.~Elia, et al., ALICE Internal Note ALICE-INT-2009-021 (2009)

\bibitem{BckNote} J.F.~Grosse-Oetringhaus et al., ALICE Internal Note ALICE-INT-2009-022 (2009)

\bibitem{GEANT} R.~Brun et al., 1985 GEANT3 User Guide, CERN Data Handling Division DD/EE/841;
R.~Brun et al., 1994 CERN Program Library Long Write-up, W5013, GEANT Detector Description and Simulation Tool

\bibitem{JanFiete} J.F.~Grosse-Oetringhaus,
PhD thesis, University of M\"{u}ns\-ter, Germany (2009), CERN-THESIS-2009-033

\bibitem{QGSMcal} A.B.~Kaidalov and M.G.~Poghosyan, submitted to Eur. Phys. J.~C, arXiv:0910.2050 [hep-ph] (2009)

\bibitem{CSCtune} A.~Moraes (ATLAS Collaboration), ATLAS Note ATL-COM-PHYS-2009-119 (2009), ATLAS CSC (306) tune

\bibitem{Perugiatune} P.Z.~Skands, Multi-Parton Interaction Workshop, Perugia, Italy, 28-31 Oct 2008, arXiv:0905.3418 [hep-ph] (2009), Perugia-0 (320) tune

\bibitem{ITS_dNdEta} W.~Thome et al., Nucl.~Phys.~B \textbf{129} (1977) 365

\bibitem{UA1} UA1 Collaboration, C.~Albajar et al., Nucl. Phys.~B \textbf{335} (1990) 261

\bibitem{UA5Rep} UA5 Collaboration, G.J.~Alner et al., Phys. Rep. \textbf{154} (1987) 247

\bibitem{CDF_dNdEta} CDF Collaboration, F.~Abe et al., Phys. Rev.~D \textbf{41} (1990) 2330










\end{thebibliography}
\end{document}